\documentclass{aa}
\usepackage{graphicx}
\usepackage{amssymb}
\begin{document}
\title{The early Stages of NOVA Oph 2003 (V2573 Oph)\thanks{Based on observations made at ESO, LaSilla Chile}}

\author{S. Kimeswenger \and M.F.M. Lechner}

\offprints{S. Kimeswenger,
\\ \email{stefan.kimeswenger@uibk.ac.at}}

\institute{Institut f{\"u}r Astrophysik der Universit{\"a}t
Innsbruck, Technikerstr. 25, A-6020 Innsbruck, Austria }

   \date{Received 1 August 2003 / Accepted 12 September 2003}

\abstract{ Intermediate resolution spectroscopy of NOVA Oph 2003
 (V2573 Oph), which was first detected March 21$^{\rm th}$ 2003 but
  reported July 19$^{\rm th}$ 2003,
obtained 19$^{\rm th}$ to 23$^{\rm rd}$ of July is presented here.
The photometry during the early phases of the object is shortly
discussed. We also retrieved very accurate astrometry of the
target in this crowded field. This is needed to be able to do
further observations of the post-nova during the next years. The
inspection of the sky survey plates gives a possible progenitor
candidate and allows to derive a lower limit for the outburst
magnitude of about 10\fm0. The spectrum shows an overall expansion
of 2200~km\,s$^{-1}$ and has clearly complex outflow
substructures. The spectroscopy identifies this object as
classical nova, "Fe II" subclass.
  }

\maketitle


\section{Introduction}
NOVA Oph 2003 was discovered July 10$^{\rm th}$ 2003 as an 11\fm4
object (Takao el al. \cite{IAUC_A1}). They point out that Tabur
detected this variable on CCD images already March 21$^{\rm th}$
2003 and that {\sl "he initially dismissed the object as a
less-urgent variable star due to its long presence on his past
images"}. There is some ongoing discussion on the validity of this
detection (Kato \cite{kato_private}). This late report reminds us
on the situation of the report of the Nova V1178 Sco (Hasada et
al. \cite{IAUC_A}; Andersen \& Kimeswenger \cite{nova01}). Thus
the first spectroscopic data with higher resolution was only
obtained July 18$^{\rm th}$ (Della Valle et al. \cite{IAUC_A2})
showing the nova to be caught during its early decline. The
variable star name given to the object is V2573 Oph (Samus
\cite{IAUC_A3}).

We obtained spectra using the ESO NTT telescope at LaSilla (July
19$^{\rm th}$ to 22$^{\rm nd}$) with the multi mode instrument
EMMI mounted. Also $V$ filter images were obtained there. This
allowed a precise astrometry.

\section{Photometric Classification}
\begin{figure}[ht]
\centerline{\resizebox{7.9cm}{!}{
\includegraphics{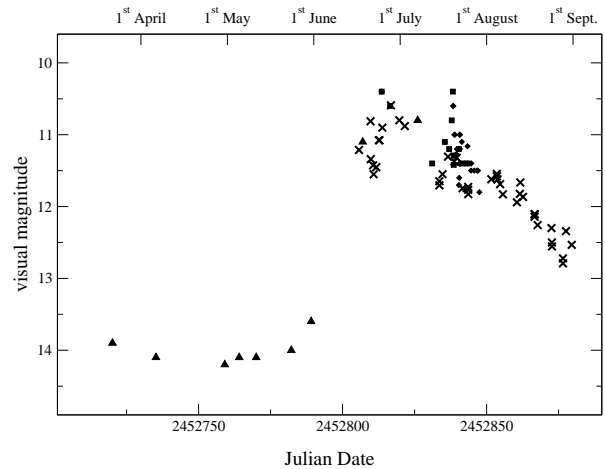}}}
 \caption{
The photometry collected from different resources; from IAUCs:
triangles - V. Tabur (possibly spurious detections - see text),
diamonds - AAVSO, squares - others in IAUCs; crosses: ASAS-3. The
R band CCD values of Tabur were shifted down by 1\fm0 to
correspond to the visual and V band measurements.} \label{phot}
\vspace{-3mm}
\end{figure}

For the light curve different resources from literature and
network were combined (Takao et al. \cite{IAUC_A1}; Liller et al.
\cite{IAUC_B2}; http://archive.princeton.edu/$\sim$asas). The data
of Tabur, labelled as unfiltered CCD images in the IAUC seem to be
taken in fact with an R band filter. The photometry before May
10$^{\rm th}$ are suspected to be spurious detections (Kato
\cite{kato_private}). It is beyond the scope and the possibilities
of the authors to verify this here. They were shifted by 1\fm0 to
fit the visual light curve and the upper limits given by Liller et
al. (\cite{IAUC_B2}) for June. This might be a major source of
uncertainty, as the target surely will have changed its color
during the early phases. The resulting light curve is shown in
Fig.~\ref{phot}. The shape is similar to the ones of V1548~Aql and
V723~Cas, which are joined to a subclass by Kato \& Takamizawa
(\cite{IBVS5100}). Although the increase from the plateau phase to
the final peak of about 3\fm0 is double that of these two older
novae, they seem to have a lot in common. Using the decline phase
data from the maximum (JD = 2452816) in the ASAS-3 data only, we
derive a $t_2 \approx 62$~days. This is very similar to that of
the slow FeII class novae like DQ~Her and V868~Cen with 67 and 55
days respectively (Della Valle \& Livio \cite{DV98}).

\section{Astrometry and Cross--identification}

To obtain a very accurate astrometry of the target a set of $V$
band images were taken. The EMMI red arm is currently equipped
with a mosaic of two 2k$\times$4k CCDs giving, according to the
manual, a resolution of 0\farcs333 per pixel. In fact we measure
0\farcs33459. The FWHM on the images used for the astrometry were
within the range of 0\farcs68 to 0\farcs81. Only the chip covering
the optical center of the field of view was used to avoid
additional free parameters like rotation between the chips or
different scales due to inclination. We used only the central
distortion free part of the image, 2\farcm7$\times$3\farcm5 in
size around the target. As the target was taken near the zenith
there should be no differential refraction. Astrometric
calibrators were taken from {\it USNO CCD Astrometric Catalogue}
(UCAC) (Zacharias et al. \cite{UCAC}). 13 stars surrounding the
target were used to obtain the astrometry. The next nearby TYCHO-2
source has a distance of more than 6\farcm0 and is known to have a
high proper motion. Thus an additional TYCHO reference system, as
in Andersen \& Kimeswenger (\cite{nova01}), was not applied here.

\begin{figure}[ht]
\resizebox{\hsize}{!}{\phantom{XXXXXXXXXXXXXXXXXXXX}
\includegraphics{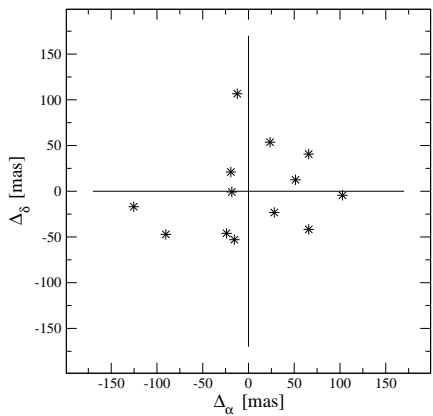}\phantom{XXXXXXXXXXXXXXXXXXXX}}
\caption{The scatter diagram of the astrometric calibration
sources around the target.} \label{astro_fig} \vspace{-3mm}
\end{figure}

The source extraction was obtained by using SExtractor v2.1.6
(Bertin \& Arnouts \cite{Bertin}). The rms of the positions was
31~mas (Fig.~\ref{astro_fig}). The largest residuals are found for
the faintest sources. As our S/N was very high even for those
sources, we assume that part of this error originates from the
UCAC. As the target is very bright, the internal accuracy of the
target coordinates was even better (5~mas rms). Assuming some
systematic effects an error estimate of 20~mas is very
conservative:

\smallskip
\centerline{~$\alpha_{\rm \tt J2000.0} = \,\,\,\,17^{\rm \tt
h}19^{\rm \tt m}14\fs0913\, \pm 0\fs0014$}
\centerline{\,$\delta_{\rm \tt J2000.0} =
-27\degr22'35\farcs315\,\,\, \pm 0\farcs020$}

\begin{figure}[ht]
\centerline{\resizebox{!}{4cm}{
\includegraphics{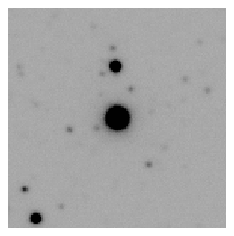}}\phantom{XX}\resizebox{!}{4cm}
{\includegraphics{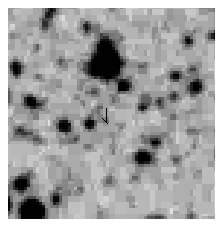}} } \caption{The finding chart for
the target from our ESO NTT observations (left) and from
SuperCOSMOS scans of the UKIST B plate. The coordinates from the
astrometry are centered (fineline $+$). There is a faint
($B\geq20\fm5$, $R>19\fm5$) star ($\times$) which might be the
progenitor (see text).} \label{image} \vspace{-3mm}
\end{figure}

These coordinates are more accurate than those given by McNaught
\& Garradd (\cite{IAUC_B1}). Their larger error bars do overlap to
the results here. An inspection of the sky survey plates SERC V
and 2$^{\rm nd}$~ed.~ER shows a source at the plate limits very
near to this position. The SuperCOSMOS scans (Hambly et al.
\cite{SuperCOSMOS}) give an estimate of B$\geq$20\fm5 and
R$>$19\fm5 (Fig.~\ref{image}) when using the PSF fitting for
deblending and photometric calibration as described in Kimeswenger
\& Weinberger (\cite{KW01}). There is no brighter source in the
field which is a candidate for the progenitor. This allows to
derive a lower limit  for the amplitude of the outburst of about
10\fm0.

\section{Spectroscopy}

\begin{figure*}[ht]
\resizebox{\hsize}{!}{\phantom{XXXX}\includegraphics{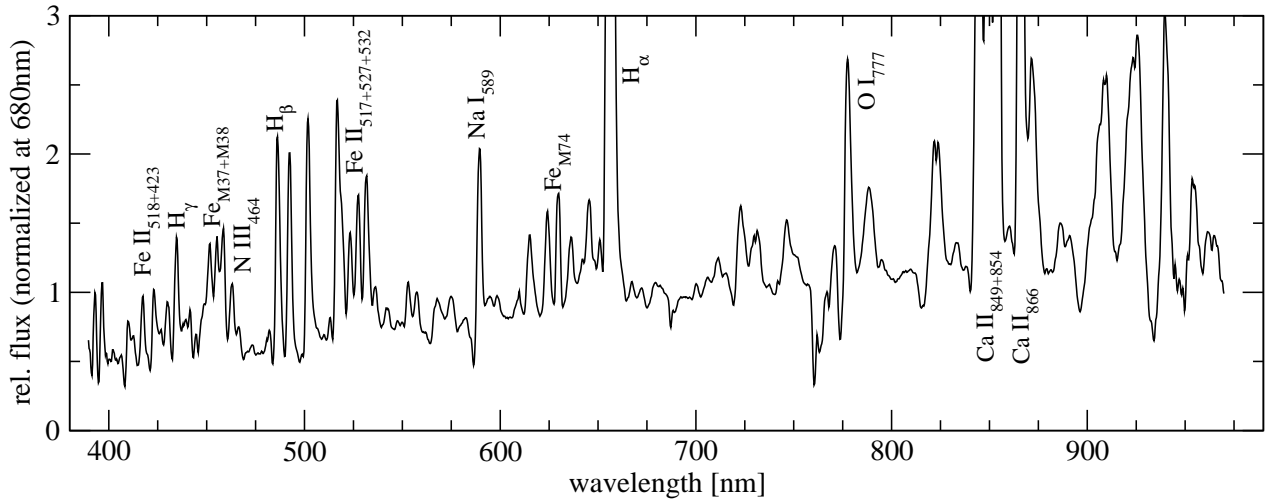}\phantom{XXXX}}
  \caption{The overall spectrum of V2573 Oph.
H$_{\alpha}$ and CaII$_{849+854+866}$ are displaced exceeding the
box in order to see all the weaker features. All P-Cygni profiles
show more or less the same global expansion, but with different
substructures. The maximum expansion velocity is about 2200
km\,s$^{-1}$. Most of the lines at the red end are blends of OI,
NI and metal lines. They are thus not labelled individually.}
\label{full_spectrum}
\end{figure*}

The spectra were obtained at the ESO NTT + EMMI. We used grism \#2
(resolution of 0.35~nm/pixel; 390 to 980 nm) and grism\#6
(resolution of 0.14~nm/pix; 500 to 880 nm). There were taken at
least two spectra with each grism during four nights, starting
19./20. July. The observations always were obtained between 23:20
and 0:40 UT. The calibration (bias, flatfield, wavelength
calibration and response curve) was done using usual procedures in
MIDAS.

\begin{figure}[h]
\resizebox{\hsize}{!}{ \phantom{XX}
\includegraphics{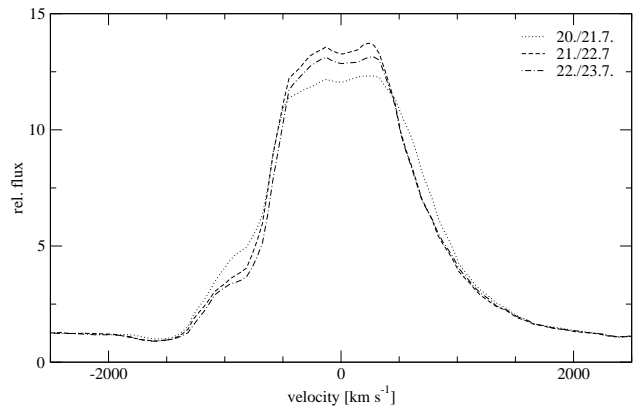} \phantom{XX} }
 \caption{The P-Cygni profile of the Hydrogen Balmer line.
 The line peak grows significantly from day to day while it gets smaller.
 Especially the emission wing around -1000~km\,s$^{-1}$ vanishes more and more.} \label{balmer} \vspace{-3mm}
\end{figure}

\begin{figure}[h]
\resizebox{\hsize}{!}{ \phantom{XX}
\includegraphics{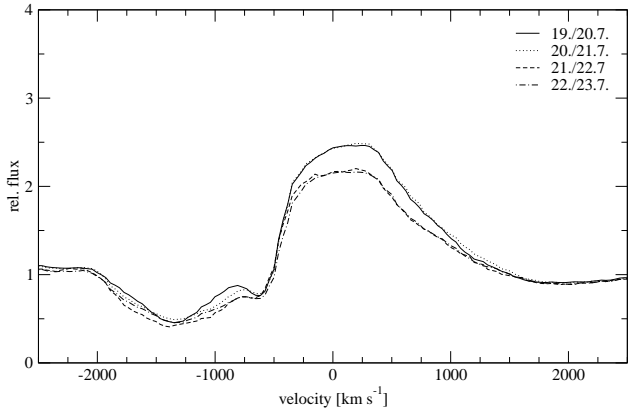}
\phantom{XX} } \caption{The profile of the OI line at 777\,nm
shows the highest expansion (see text).} \label{OI} \vspace{-3mm}
\end{figure}

\begin{figure}[h]
\resizebox{\hsize}{!}{ \phantom{XX}
\includegraphics{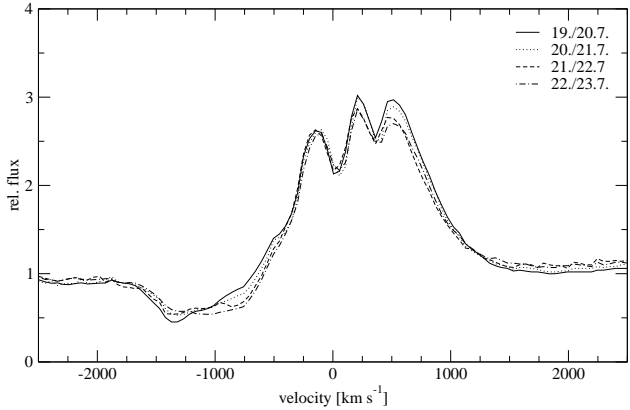}
\phantom{XX} } \caption{The profile of the Na-D line shows nearly
the same profile like the Balmer lines. The emission feature at
-1000~km\,s$^{-1}$ also faded during the observation period.}
\label{NaI} \vspace{-3mm}
\end{figure}

The spectra had a S/N of $>$200 in the continuum over the whole
region (Fig.~\ref{full_spectrum}). The intensity of most of the lines
slowly decreased (Tab.~\ref{lines_tab})
during the observation period.
In the night 19./20. July the H$_\alpha$ and the Ca IR triplet were overexposed.\\
Lines of the Balmer series (Fig.~\ref{balmer}), OI (Fig.~\ref{OI})
and NaI (Fig.~\ref{NaI}) show significant P-Cygni profiles giving
expansions of up to 1800, 2200 and 1600 km\,s$^{-1}$ respectively.
This is somewhat higher but within the same range than the values
measured July 18$^{th}$ by Della\,Valle et al. (\cite{IAUC_A2}).
Remarkably, the iron lines are stronger than e.g. reported in CI
Aql (Kiss et al. \cite{Kiss}), while [NII]$_{575}$ hardly is
visible. The Na-D line is very prominent too. Normally blended by
HeI$_{588}$, this line appears here only as a weak absorption
feature at the blue end. This part of the spectrum is very much
like that of Nova Sco 2001 (Andersen \& Kimeswenger
\cite{nova01}).

\begin{table}[ht]
\caption{Observed emission line fluxes (unit:
10$^{-14}$\,W\,m$^{-2}$\,$\mu$m$^{-1}$) for the most prominent
lines which are well separated (no blends). The fluxes were
absolute calibrated with respect to the ASAS-3 photometry ($m_V =
11\fm34$) by folding standard filter curves. The main uncertainty
originates from the 'definition' of the continuum around the
lines.} \label{lines_tab}
\begin{tabular}{lcccc}
\hline
Line  &  July, 19$^{\rm th}$ &  20$^{\rm th}$&  21$^{\rm th}$ &  22$^{\rm nd}$ \\
\hline \hline
H$_\alpha$      & {\tt ~2.1} & {\tt ~2.1} & {\tt ~2.0} & {\tt ~1.8}\\
H$_\beta$       & {\tt ~6.3} & {\tt ~5.9} & {\tt ~5.5} & {\tt ~5.5}\\
H$_\gamma$      & overexp.   & {\tt 58.5} & {\tt 66.0} & {\tt 63.4}\\
  &  \\
Fe II 418 nm    & {\tt ~1.8} & {\tt ~1.7} & {\tt ~1.5} & {\tt ~1.6} \\
Fe II 423 nm    & {\tt ~1.5} & {\tt ~1.5} & {\tt ~1.2} & {\tt ~1.1} \\
Fe II 517 nm    & {\tt ~7.7} & {\tt ~7.0} & {\tt ~7.6} & {\tt ~7.2} \\
Fe II 527 nm    & {\tt ~2.6} & {\tt ~2.2} & {\tt ~2.5} & {\tt ~2.5} \\
Fe II 532 nm    & {\tt ~3.7} & {\tt ~3.2} & {\tt ~3.7} & {\tt ~3.4} \\
 &  &  &  &  \\
N II 464nm      & {\tt ~1.6} & {\tt ~1.6} & {\tt ~1.6} & {\tt ~1.5} \\
  &  \\
Na I 589nm      & {\tt ~4.6} & {\tt ~4.0} & {\tt ~3.7} & {\tt ~3.7} \\
  &  \\
O I 777nm       & {\tt ~7.5} & {\tt ~7.7} & {\tt ~7.7} & {\tt ~7.4} \\
  &  \\
Ca II 849+854nm & overexp.   & {\tt 56.3} & {\tt 54.6} & {\tt 48.9} \\
Ca II 866 nm    & overexp.   & {\tt 27.3} & {\tt 27.3} & {\tt 24.6} \\
\hline \hline
\end{tabular}
\end{table}

\begin{figure}[h]
\resizebox{\hsize}{!}{ \phantom{XX}
\includegraphics{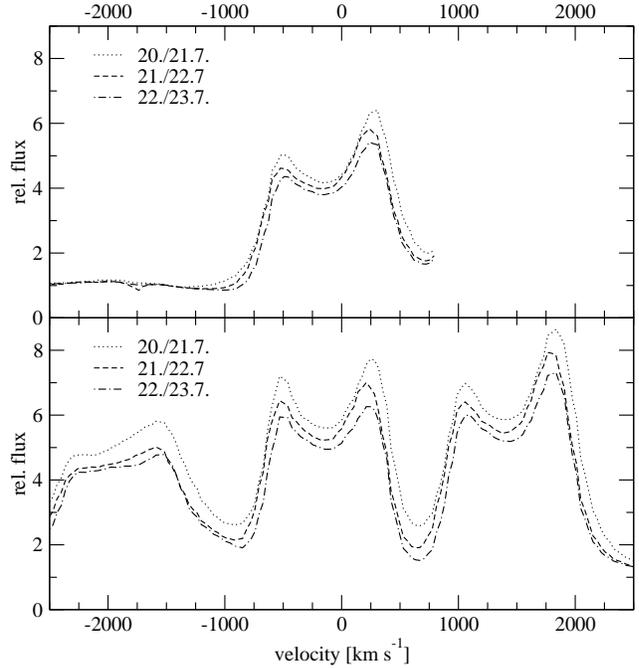}
\phantom{XX} } \caption{The profile of the IR CA II lines at
866\,nm (upper panel) and at 849+854\,nm (lower panel). They show
now P-Cygni type absorption but a nicely pronounced optically thin
shell structure with an expansion of 550~km\,s$^{-1}$. The
structure is blueshifted by about 150~km\,s$^{-1}$.} \label{CaII}
\vspace{-3mm}
\end{figure}

The Ca II IR triplet does not show P-Cygni profiles, but a
pronounced saddle like profile (Fig.~\ref{CaII}). This may
originate from a tilted equatorial expanding ring or bipolar
ejected clumps. Although assuming spheroidicity, this also can be
explained by a transition of the object from photosphere to shell
type spectra as shown in Williams (\cite{will92}) for Nova~LMC
1988~No.~2 for the hydrogen lines. As Ca~II has the lowest
excitation of the optical lines here, the object was maybe just
starting this process at its outer cooler parts.
The structure seems to be
blueshifted with respect to the rest wavelength of the line by
about 150~km\,s$^{-1}$.

\section{Conclusion}

The spectrum classifies this object as of "Fe II" subtype (after
Williams \cite{will92}). The astrometry presented here clearly
indicates that the possible progenitors are at the plate limit of
the sky surveys or beyond. This allows us to give a lower limit
for the outburst magnitude of $\Delta M \geq 10\fm0$. The
photometry classifies it as slow nova, with some signature of
starting humps during decline as e.g. described in Kato et al.
(\cite{IBVS5309}). On the other hand the outburst magnitude is
unusual high for a slow nova (Kato et al. \cite{IBVS5309}). Also
the derived expansion is high for a typical slow nova: e.g.
V356~Aql had 450~km\,s$^{-1}$ (McLaughlin \cite{V356}) and RR~Tel
had not more than 600~km\,s$^{-1}$ (Thackeray \& Webster
\cite{RRtel}). Thus the FeII-broad classification (Williams
\cite{will92}) may be applied here. The pre max halt 3\fm5 below
maximum is very unusual at for a $t_2 \approx 62^d$.

Della Valle (\cite{DV02}) give $M_V = 7\fm2^{+1\fm2}_{-0\fm9}$ for
such kind of slow novae. As there is no information about the
interstellar reddening we estimate with the information of
DellaValle et al. (\cite{IAUC_A2}) an extinction of $A_V < 5\fm0$.
This leads to a crude estimate for the distance of 3.5-4.0 kpc.
Future measurements (namely the expansion parallax of the
post-nova shell) are needed to fix the distance and thus the
absolute magnitude of the object.
\\
The spectroscopy shows a complex structure of the outflow.   The
velocity field obtained by the hydrogen and helium lines suggests
a two shell structure similar to the models for Supernova 1997A
(Hanuschik et al. \cite{shell}) or strong aspherical components
like expanding rings or bipolar outflows. The metal lines even
indicate a more complex, most likely non--symmetric outflow with
respect to the line of sight.

\begin{acknowledgements}
We thank the referee M. Della Valle for his helpful suggestions.
KS is grateful to the {\it Bun\-des\-mini\-sterium f\"ur
Bil\-dung, Wis\-sen\-schaft und Kultur} (BMfBWK) for travel
support.
\end{acknowledgements}


\begin{thebibliography}{}
\bibitem[2001]{nova01}
Andersen, M., \& Kimeswenger, S. 2001, A\&A, 377, L5

\bibitem[1996]{Bertin}
Bertin, E., \& Arnouts, S. 1996, A\&AS, 117, 393

\bibitem[2002]{DV02}
Della Valle, M. 2002, in Classical Nova Explosions. AIP Conference
Proceedings, Vol. 637. Eds M. Hernanz \& J. Jos{\'e}. American
Institute of Physics, p.443

\bibitem[1998]{DV98}
Della Valle, M., \& Livio, M. 1998, ApJ, 506, 818

\bibitem[2003]{IAUC_A2}
Della Valle, M., Mason, E., Pasquini, L., \& Prichard, J. 2003,
IAUC, 8166, 2


\bibitem[2001]{SuperCOSMOS}
Hambly, N.C., Irwin, M.J.,  \& MacGillivray, H.T. 2001, MNRAS,
326, 1295

\bibitem[1993]{shell}
Hanuschik, R.W, Spyromilio, S., Stathakis, R., Kimeswenger, S.,
Gochermann, J., Seidensticker, K.J., \& Meurer, G. 1993, MNRAS,
261, 909

\bibitem[2001]{IAUC_A}
Haseda, K., Kadota, K., Yamaoka, H., Takamizawa, K., \& Kato, T.
2001, IAUC, 7647, 1

\bibitem[2003]{kato_private}
Kato, T. 2003, September, 21$^{\rm th}$, private communication

\bibitem[2001]{IBVS5100}
Kato, T., \& Takamizawa, K. 2001, IBVS, 5100, 1

\bibitem[2002]{IBVS5309}
Kato, T., Yamaoka, H., \& Ishioka, R. 2002, IBVS, 5309, 1

\bibitem[2001]{KW01}
Kimeswenger, S., \& Weinberger, R. 2001, A\&A, 370, 991


\bibitem[2001]{Kiss}
Kiss, L.L., Thomson, J.R., Ogloza, W., F{\"u}r{\'e}sz, G., \&
Szil{\'a}di, K. 2001, A\&A, 366, 858

\bibitem[2003]{IAUC_B2}
Liller, W., West, J.D., Brown, N.J., Linnolt, M., Lehky, M.,
Baransky, A., \& Carvajal, J. 2003, IAUC, 8167, 2

\bibitem[1955]{V356}
McLaughlin, D.B. 1955, ApJ 122, 417

\bibitem[2003]{IAUC_B1}
McNaught, R.H., \& Garradd, G.J. 2003, IAUC, 8167, 1


\bibitem[2003]{IAUC_A3}
Samus, N.N. 2003, IAUC, 8166, 3



\bibitem[2003]{IAUC_A1}
Takao, A., Monard, L.A.G., Tabur, V., \& Kato, T. 2003, IAUC,
8166, 1

\bibitem[1974]{RRtel}
Thackeray, A.D., \& Webster, B.L. 1974, MNRAS, 168, 101





5 Feast, M., 2000, MNRAS, 319, 728

\bibitem[1992]{will92}
Williams, R.E., 1992, AJ, 104, 725

\bibitem[2000]{UCAC}
Zacharias, N., Urban, S.E., Zacharias, M.I., et al., 2000, AJ, 120, 2131
\end{thebibliography}
\end{document}